\documentclass[aps,prb,eqsecnum,superscriptaddress,amsmath,twocolumn]{revtex4}

\usepackage{graphicx}

\usepackage{bm}
\def\kv{\mbox{\boldmath $ k$}}

\begin{document}
\preprint{Preprint}

\title{
Nuclear magnetic relaxation and superfluid density in Fe-pnictide superconductors: 
An anisotropic $\pm s$-wave scenario
}

\author{Yuki Nagai}
\affiliation{
Department of Physics, University of Tokyo, Tokyo 113-0033, Japan
}

\author{Nobuhiko Hayashi}
\affiliation{
CCSE, Japan  Atomic Energy Agency, 6-9-3 Higashi-Ueno, Tokyo 110-0015, Japan
}
\affiliation{
CREST(JST), 4-1-8 Honcho, Kawaguchi, Saitama 332-0012, Japan
}

\author{Noriyuki Nakai}
\affiliation{
CCSE, Japan  Atomic Energy Agency, 6-9-3 Higashi-Ueno, Tokyo 110-0015, Japan
}
\affiliation{
CREST(JST), 4-1-8 Honcho, Kawaguchi, Saitama 332-0012, Japan
}

\author{Hiroki~Nakamura}
\affiliation{
CCSE, Japan  Atomic Energy Agency, 6-9-3 Higashi-Ueno, Tokyo 110-0015, Japan
}

\author{Masahiko Okumura}
\affiliation{
CCSE, Japan  Atomic Energy Agency, 6-9-3 Higashi-Ueno, Tokyo 110-0015, Japan
}
\affiliation{
CREST(JST), 4-1-8 Honcho, Kawaguchi, Saitama 332-0012, Japan
}

\author{Masahiko Machida}
\affiliation{
CCSE, Japan  Atomic Energy Agency, 6-9-3 Higashi-Ueno, Tokyo 110-0015, Japan
}
\affiliation{
CREST(JST), 4-1-8 Honcho, Kawaguchi, Saitama 332-0012, Japan
}

\date{\today}

\begin{abstract}
We discuss the nuclear magnetic relaxation rate and the superfluid density with the use of 
the effective five-band model by
Kuroki {\it et al}.\ [Phys.\ Rev.\ Lett.\ {\bf 101}, 087004 (2008)]
in Fe-based superconductors.
We show that a fully-gapped anisotropic $\pm s$-wave superconductivity 
consistently explains experimental observations.
In our phenomenological model,
the gaps are assumed to be 
anisotropic on the electron-like $\beta$ Fermi surfaces around the $M$ point,
where
the maximum of the anisotropic gap is about four times larger than the minimum.
\end{abstract}


\maketitle

\section{Introduction}
Much attention has been focused on
novel Fe-based superconductors
since the recent discovery of superconductivity at the high temperature 26 K
in LaFeAsO$_{1-x}$F$_x$.\cite{Kamihara}
Up to now, many Fe-based superconductors (especially iron pnictides)
such as SmFeAsO$_{1-x}$F$_x$
have been found and intensively
investigated.\cite{Kamihara06P,Watanabe07Ni,Takahashi,Chen08Ce,Ren08Pr,Ren08Nd,Kito08Nd,Ren08Sm,Chen08Sm,Yang08Gd,Wang08Gd,Fang08LaSrNi,Rotter08BaK,Takeshita08NdwoFpress,Ren08woF,Hsu08FeSe}
Experimental observations of thermodynamic quantities and others 
begin now to be reported on those 
superconductors.\cite{mu,sefat,kohama,dong,ren,luetkens,ahilan,khasanov,malone,hashimoto,luetkens2,takeshita,martin,nakai,grafe,matano,mukuda,kotegawa} 
Recently, 
the superfluid density 
and the nuclear spin-lattice relaxation rate have been analyzed theoretically.\cite{parker,chubukov,bang0807,bang,benfatto} 
Such observations and analyses are important and indispensable for elucidating superconducting properties,
especially for Cooper-pairing symmetry which we will discuss.

One of the confused points in the experiments for Fe-based superconductors 
is that the results of the nuclear magnetic relaxation rate seem inconsistent with 
the superfluid density observations. 
The nuclear magnetic relaxation rate has the lack of the coherence peak below $T_{\rm c}$
and exhibits the low temperature 
power-law behavior ($1/T_{1} \propto T^{3}$).\cite{nakai,grafe,matano,mukuda,kotegawa} 
This is seemingly the evidence of unconventional superconductivity with line-node gaps. 
However, 
some experiments report that 
the superfluid density (i.e., penetration depth) does not depend on the temperature at low temperatures, which means 
that the pairing symmetry is fully-gapped $s$-wave symmetry.\cite{luetkens,khasanov,malone,hashimoto,martin,luetkens2}
The $\pm s$-wave pairing symmetry 
is theoretically proposed 
as one of the candidates for the pairing symmetry
in Fe-pnictide superconductors.\cite{bang0807,parish,kuroki,seo,Arita,mazin,eremin,nomura,stanev,senga}
The $\pm s$-wave symmetry means that
the symmetry of pair functions on each Fermi surface is $s$-wave
and the relative phase between them is $\pi$.
Very recently, several theoretical groups suggested that the $\pm$s-wave symmetry explains the lack of the coherence peak 
and the low temperature power-law behavior in 
the nuclear magnetic relaxation rate, with introducing impurity scatterings.\cite{parker,chubukov,bang}
Part of their scenarios is based on the fact that,
in a $\pm s$-wave phase,
substantial low-energy states appear in the density of states
in the case of a unitary-limit scattering,
while only higher-energy density of states near gap edges is modified
when approaching to the Born limit.\cite{bang,preosti}

To theoretically investigate the superconductivity, it is necessary to consider a model
for the electronic structure.
There are many theoretical studies, especially by band calculations, to understand the unique 
electronic and magnetic properties of those Fe-pnictide superconductors. \cite{lebegue,singh,cao,haule,xu-band,ishibashi,boeri,yin,yildirim,shorikov,nakamura,kuroki,knakamura,nekrasov,hli,sushko,craco,ma0806,cvetkovic,wu0805} 
In addition, 
an effective five-band model was elaborated by Kuroki {\it et al.},\cite{kuroki}
where the five bands originate predominantly from $3d$ orbitals at the Fe atomic site.
A simpler two-band Hamiltonian was also proposed
as a tractable minimal model,
which reproduces the structure of Fermi surfaces obtained
by band calculations.\cite{han,li,raghu,graser,daghofer}
However, Arita {\it et al}.\cite{Arita} claimed that the five bands are necessary for describing correct band dispersions around the Fermi level.
They also suggested that an {\it anisotropic} $\pm s$-wave superconductivity is realized in a Fe-pnictide superconductor.\cite{Arita}

In this paper, 
we investigate the nuclear spin-lattice relaxation rate and the superfluid density 
on the basis of the realistic effective five-band model. 
We will show that an anisotropic $\pm s$-wave 
pair function explains consistently the experimental results even in assuming a rather clean system.

This paper is organized as follows.
The effective five-band model and the pair functions are introduced in Sec.\ II.
We then discuss the nuclear spin-lattice relaxation rate (Sec.\ III), 
the superfluid density (Sec.\ IV), and the density of states (Sec.\ V). 
Finally, the conclusion is given in Sec.\ VI.
   In the appendix,
we describe the derivation of the nuclear spin-lattice relaxation rate
on the basis of the quasiclassical theory of superconductivity.

\section{Model}
\label{sec:model}
\begin{figure}
  \begin{center}
    \begin{tabular}{p{40mm}p{40mm}}
      \resizebox{40mm}{!}{\includegraphics{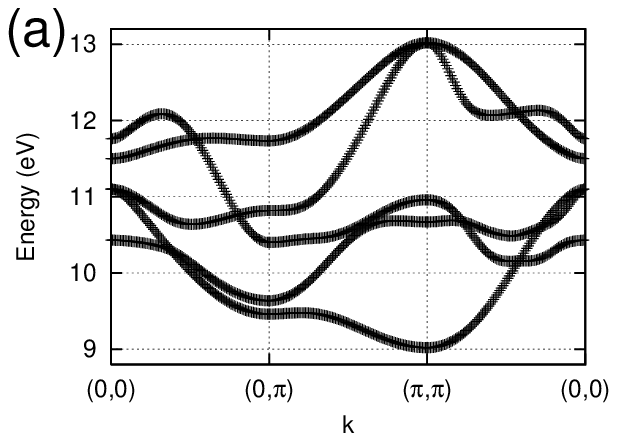}} &
      \resizebox{40mm}{!}{\includegraphics{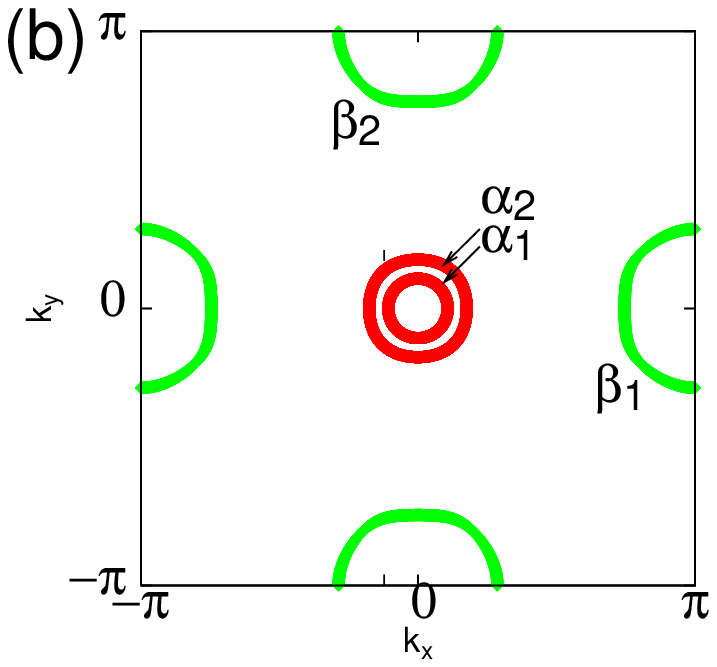}} 
    \end{tabular}
\caption{\label{fig:fig0}
(Color online) (a) Band dispersion of the effective five-band model and 
(b) Fermi surfaces with the Fermi energy $E_{\rm F} = 10.97$eV.
}
  \end{center}
\end{figure}

We introduce the effective five-band model proposed by Kuroki {\it et al}.\cite{kuroki} 
The tight-binding Hamiltonian 
is written as
\begin{eqnarray}
H_{0} &=& \sum_{ij} \sum_{\mu \nu} \sum_{\sigma} \Bigl{[} 
t(x_{i} - x_{j},y_{i}-y_{j};\mu,\nu) c_{i \mu \sigma}^{\dagger} c_{j \nu \sigma} \nonumber \\
& &+t(x_{j} - x_{i},y_{j} - y_{i};\nu,\mu) c^{\dagger}_{j \nu \sigma} c_{i \mu \sigma}
\Bigl{]}
+ \sum_{i \mu \sigma} \epsilon_{\mu} n_{i \mu \sigma}, \nonumber \\ 
\end{eqnarray}
where $c^{\dagger}_{i \mu \sigma}$ creates an electron with spin $\sigma$ on the $\mu$-th orbital at site $i$, 
$n_{i \nu \sigma} = c^{\dagger}_{i \mu \sigma} c_{i \mu \sigma}$, and $t$ denotes the hopping parameters. 
Here, the onsite energies are $(\epsilon_{1},\epsilon_{2},\epsilon_{3},\epsilon_{4},\epsilon_{5}) 
= (10.75,10.96,10.96,11.12,10.62)$eV and 
the hopping parameters are considered 
up to fifth nearest neighbors 
(see a table in Ref.\ \onlinecite{kuroki}). 
The band dispersion of this model is shown in Fig.~\ref{fig:fig0}(a), and 
the Fermi surfaces are shown in Fig.~\ref{fig:fig0}(b). 
There are two hole pockets (denoted as $\alpha_{1}$, $\alpha_{2}$) centered around 
$(k_{x},k_{y}) = (0,0)$ and two electron pockets around 
$(\pi,0) (\beta_{1})$ or $(0,\pi)(\beta_{2})$. 

\begin{figure}
\begin{center}
\includegraphics[width = 7cm]{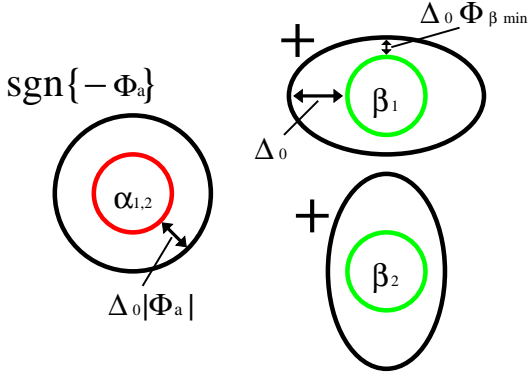}
\caption{\label{fig:fig0-2}
(Color online) Schematic figures of the pair functions on each Fermi surface.
}
  \end{center}
\end{figure}

Arita {\it et al}.\cite{Arita} have performed five-band RPA calculations for the five-band model. 
Their result 
suggests that 
the pairing function is an anisotropic $\pm s$-wave symmetry. 
This pairing has 
isotropic $s$-wave pair functions on the Fermi surfaces $\alpha_{1,2}$, and 
anisotropic $s$-wave pair functions 
on $\beta_{1,2}$ where 
the maximum of the pair amplitude is about five times larger than 
the minimum.\cite{Arita} 
Following it, 
we assume phenomenologically
the anisotropic $\pm s$-wave pair function expressed as
(see Fig.\ \ref{fig:fig0-2})
\begin{eqnarray}
\Delta_{\alpha_{1,2},\beta_{1,2}}(\kv) &=& \Delta_0 \Phi_{\alpha_{1,2},\beta_{1,2}}(\kv) \tanh(a \sqrt{T_c/T-1}), 
\: \: \: \: \: \: \: \:  \label{eq:tem}\\
\Phi_{\alpha_{1,2}}(\kv) &=& - \Phi_a,\\
\Phi_{\beta_{1,2}}(\kv) &=& \frac{(1 + \Phi_{\beta {\rm min}} )}{2} \pm \frac{(1- \Phi_{\beta {\rm min}} ) \cos (2 \phi_{1,2})}{2}.  \nonumber \\
\end{eqnarray}
Here, 
$\Phi_{\alpha_{1(2)}}(\kv)$ and $\Phi_{\beta_{1(2)}}(\kv)$ denote the pair amplitudes 
on the Fermi surfaces $\alpha_1$($\alpha_2$) and $\beta_1$($\beta_2$), respectively. 
Equation 
(\ref{eq:tem}) with $a = 1.74$ reproduces well 
the temperature dependence of the BCS gap. 
The angles $\phi_{1}$ and $\phi_{2}$ are measured from 
the $(\pi, 0)$ direction around 
$(k_x,k_y) = (\pi,0)$ and $(0,\pi)$, respectively.
The range of the gap-anisotropy parameter $\Phi_{\beta {\rm min}}$ is
$0 \le \Phi_{\beta {\rm min}} \le 1$.
The larger $\Phi_{\beta {\rm min}}$ within this range, the weaker anisotropy.
The sign of $-\Phi_a$ corresponds to the relative phase of the pair functions
between the $\alpha$ and $\beta$ Fermi surfaces.
If $\Phi_a$ is positive (negative), the pairing
is $\pm s$-wave ($s$-wave).

The pair functions on the Fermi surfaces $\alpha_{1,2}$ are isotropic and 
those on $\beta_{1,2}$ are anisotropic.
The isotropic gap amplitude on $\alpha_{1,2}$ is $\Delta_0 |\Phi_a|$.
The anisotropic gaps on $\beta_{1,2}$
have the maximum (minimum) value $\Delta_{0}$ ($\Delta_{0} \Phi_{\beta {\rm min}}$).
From the RPA results presented in Ref.\ \onlinecite{Arita}, 
it seems that 
$\Phi_a \sim 0.2$ and 
$\Phi_{\beta {\rm min}} \sim 0.2$. 
With 
adjusting 
$\Phi_a$, $\Phi_{\beta {\rm min}}$, and $\Delta_0/T_c$ as parameters, 
we will calculate the nuclear magnetic relaxation rate $1/T_1$ and the superfluid density $\rho_{xx}$. 
We consider the following pair functions: 
(i) isotropic $s$-wave ($\Phi_{a} < 0$, $\Phi_{\beta {\rm min}} = 1$), (ii) anisotropic $s$-wave ($\Phi_{a} < 0$, $\Phi_{\beta {\rm min}} \neq 1$), (iii) isotropic $\pm s$-wave ($\Phi_{a} > 0$, $\Phi_{\beta {\rm min}} = 1$), and (iv) anisotropic $\pm s$-wave 
($\Phi_{a} > 0$, $\Phi_{\beta {\rm min}} \neq 1$).
Here, we exclude spin-triplet pairings because
Knight-shift measurements suggest a spin-singlet pairing.\cite{grafe,matano}

\section{Nuclear spin-lattice relaxation rate}
\label{sec:nmr}
The nuclear spin-lattice relaxation rate 
$1/T_1 T$ is given as \cite{Haya06-2,Haya03,Haya06-B,Haya03-JLTP}
(see Appendix)
\begin{eqnarray}
\frac{T_1(T_{\mathrm c})T_{\mathrm c}}{T_1(T)T}
&=&
\frac{1}{4T }
\int_{-\infty}^{\infty} 
\frac{d\omega}{\cosh^2(\omega/2T)} W(\omega),
\end{eqnarray}
with 
\begin{eqnarray}
W(\omega) &=& \bigl\langle a^{22}_{\downarrow\downarrow}(\omega) \bigr\rangle_{\rm FS}
\bigl\langle a^{11}_{\uparrow\uparrow}(-\omega) \bigr\rangle_{\rm FS} 
 - 
\bigl\langle a^{21}_{\downarrow\uparrow}(\omega) \bigr\rangle_{\rm FS}
\bigl\langle a^{12}_{\uparrow\downarrow}(-\omega) \bigr\rangle_{\rm FS}. \nonumber \\ \\
&\equiv& W_{GG}(\omega) + W_{FF}(\omega).
\label{eq:t1t}
\end{eqnarray}
Here,
\begin{eqnarray}
a^{11}_{\uparrow\uparrow}({\bm k}_{\rm F},\omega)
&=&
\frac{1}{2}
\Bigl[
g_{\uparrow\uparrow}({\bm k}_{\rm F},i\omega_n\rightarrow \omega+ i\eta) \nonumber \\
& & -
g_{\uparrow\uparrow}({\bm k}_{\rm F},i\omega_n\rightarrow \omega- i\eta)
\Bigr],
\\
a^{22}_{\downarrow\downarrow}({\bm k}_{\rm F},\omega)
&=&
\frac{1}{2}
\Bigl[
{\bar g}_{\downarrow\downarrow}({\bm k}_{\rm F},i\omega_n\rightarrow \omega+ i\eta) \nonumber \\
 & & -
{\bar g}_{\downarrow\downarrow}({\bm k}_{\rm F},i\omega_n\rightarrow \omega- i\eta)
\Bigr],
\\
a^{12}_{\uparrow\downarrow}({\bm k}_{\rm F},\omega)
&=&
\frac{i}{2}
\Bigl[
f_{\uparrow\downarrow}({\bm k}_{\rm F},i\omega_n\rightarrow \omega+ i\eta) \nonumber \\
& & -
f_{\uparrow\downarrow}({\bm k}_{\rm F},i\omega_n\rightarrow \omega- i\eta)
\Bigr],
\\
a^{21}_{\downarrow\uparrow}({\bm k}_{\rm F},\omega)
&=&
\frac{i}{2}
\Bigl[
{\bar f}_{\downarrow\uparrow}({\bm k}_{\rm F},i\omega_n\rightarrow \omega+ i\eta) \nonumber \\
& & -
{\bar f}_{\downarrow\uparrow}({\bm k}_{\rm F},i\omega_n\rightarrow \omega- i\eta)
\Bigr],
\end{eqnarray}
and 
\begin{eqnarray}
\label{eq:g}
g_{\uparrow\uparrow}({\bm k}_{\rm F},i\omega_n)
=
{\bar g}_{\downarrow\downarrow}({\bm k}_{\rm F},i\omega_n)
=
\frac{\omega_n}
     {\sqrt{\omega_n^2+|\Delta({\bm k}_{\rm F})|^2} },
 \\
f_{\uparrow\downarrow}({\bm k}_{\rm F},i\omega_n)
=
\frac{\Delta({\bm k}_{\rm F})}
     {\sqrt{\omega_n^2+|\Delta({\bm k}_{\rm F})|^2} },
 \\
{\bar f}_{\downarrow\uparrow}({\bm k}_{\rm F},i\omega_n)
=
\frac{\Delta^*({\bm k}_{\rm F})}
     {\sqrt{\omega_n^2+|\Delta({\bm k}_{\rm F})|^2} }.
\end{eqnarray}
The brackets $\langle \cdots \rangle_{\rm FS}$ mean the Fermi-surface average,
\begin{eqnarray}
\bigl\langle \cdots \bigr\rangle_{\rm FS}
=
\frac{
\displaystyle \sum_{i = \alpha_1,\alpha_2,\beta_1.\beta_2} \int \cdots \frac{dS_{{\rm F},i}}{\bigl|{\bm v}_{{\rm F}}({\bm k}_{\rm F})\bigr|}
}
{\displaystyle \sum_{i = \alpha_1,\alpha_2,\beta_1.\beta_2} \int \frac{dS_{{\rm F},i}}{\bigl|{\bm v}_{{\rm F}}({\bm k}_{\rm F})\bigr|}
},
\end{eqnarray}
where $dS_{{\rm F},i}$ is the area elements on 
each Fermi surface. 
$\omega_n=\pi T (2n+1)$ is the Matsubara frequency. 
We use units in which $\hbar = k_{\rm B} = 1$. 
We assume the smearing factor $\eta = 0.1 T_{\rm c}$.\cite{mean_free_path}
We set $2\Delta_0/T_{\rm c}=4$, which is a representative value near the BCS value 3.53.
The coherence factor is represented as $1+W_{FF}/W_{GG}$. 
The contribution of $W_{FF}$ is related to the coherence effect, which becomes zero in the case of 
unconventional pair functions such as $d$-wave one.

\begin{figure}
  \begin{center}
    \begin{tabular}{p{40mm}p{40mm}}
      \resizebox{55mm}{!}{\includegraphics{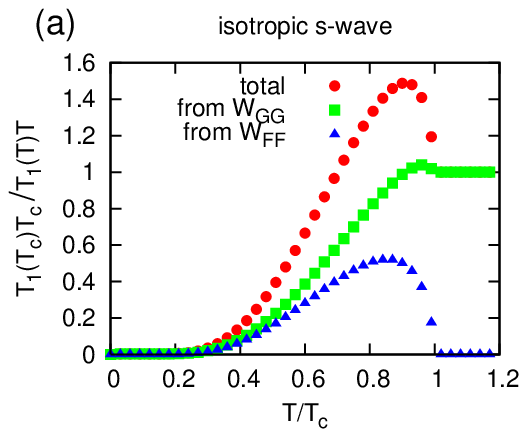}} &
      \resizebox{55mm}{!}{\includegraphics{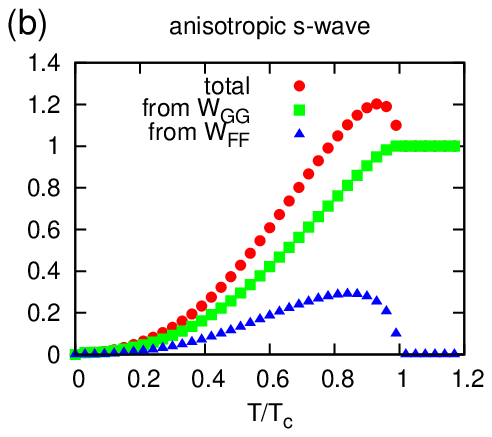}} 
    \end{tabular}
\caption{\label{fig:fig1}
(Color online) Temperature dependence of the nuclear magnetic relaxation rate $1/T_{1} T$ 
(red circles) 
with the five band 
model in the case of 
(a) the isotropic $s$-wave ($\Phi_{a} = -1$, $\Phi_{\beta {\rm min}} = 1$) and 
(b) the anisotropic $s$-wave ($\Phi_{a} = -1$, $\Phi_{\beta {\rm min}} = 0.2$). 
$2 \Delta_{0}/T_{\rm c} = 4$ and smearing factor $\eta = 0.1 T_{\rm c}$. 
The green squares denote the contribution from 
$W_{GG}$ related to the density of the states 
and the blue triangles denote 
$W_{FF}$ related to the coherence effect. 
}
  \end{center}
\end{figure}

First, we consider conventional $s$-wave pair functions ($\Phi_{a} < 0$). 
In Fig.\ \ref{fig:fig1}, we show the results for 
the isotropic $s$-wave 
($\Phi_{a} = -1$, $\Phi_{\beta {\rm min}} = 1$) 
and 
the 
anisotropic $s$-wave ($\Phi_{a} = -1$, $\Phi_{\beta {\rm min}} = 0.2$). 
The coherence peaks 
appear 
below $T_{\rm c}$ for both pair functions 
because of non-zero $W_{FF}$, meaning that 
these pair functions cannot explain the experiments.  

\begin{figure}
  \begin{center}
    \begin{tabular}{p{40mm}p{40mm}}
      \resizebox{55mm}{!}{\includegraphics{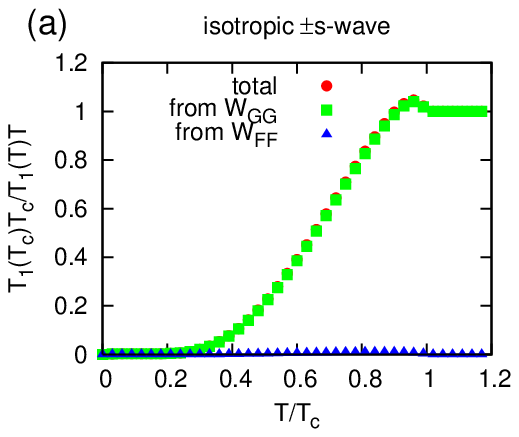}} &
      \resizebox{55mm}{!}{\includegraphics{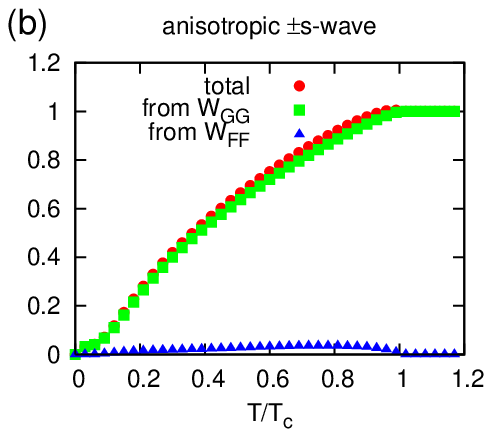}} 
    \end{tabular}
\caption{\label{fig:fig2}
(Color online)
Temperature dependence of the nuclear magnetic relaxation rate $1/T_{1}T$ (red circles) 
with the five band 
model in the case of 
(a) the isotropic $\pm s$-wave ($\Phi_{a} = 1$, $\Phi_{\beta {\rm min}} = 1$) and 
(b) the anisotropic $\pm s$-wave ($\Phi_{a} = 0.2$, $\Phi_{\beta {\rm min}} = 0.2$). 
$2 \Delta_{0}/T_{\rm c} = 4$ and smearing factor $\eta = 0.1 T_{\rm c}$.
The green squares denote the contribution from 
$W_{GG}$ related to the density of the states 
and the blue triangles denote 
$W_{FF}$ related to the coherence effect. 
}
  \end{center}
\end{figure}

Second, we consider $\pm s$-wave pair functions ($\Phi_{a} > 0$). 
Figure \ref{fig:fig2}(a) shows the result in the case of the isotropic $\pm s$-wave pair function 
($\Phi_{a} = 1$, $\Phi_{\beta {\rm min}} = 1$) and 
Fig.\ \ref{fig:fig2}(b) shows the result in the case of the anisotropic $\pm s$-wave function
($\Phi_{a} = 0.2$, $\Phi_{\beta {\rm min}}  =0.2$) 
whose $k$-dependence is similar to the result of the RPA calculation by Arita {\it et al.}\cite{Arita} 
In both cases, 
the coherence peak below $T_{\rm c}$ is suppressed, 
since $W_{FF}$ is almost zero. 
In the five band model, 
the difference of the density of states between the Fermi surfaces $\alpha_{1,2}$ and $\beta_{1,2}$ 
is small. 
Therefore, the cancellation of the $\pm s$-wave pair functions between $\alpha$ and $\beta$ 
is almost perfect, resulting in $W_{FF} \approx 0$. 
On the other hand, 
the temperature dependence at low $T$ 
is inconsistent with the experiments 
in both cases. 
The exponential behavior appears in the isotropic $\pm s$-wave case.\cite{parker} 
The temperature dependence is concave down
in the anisotropic $\pm s$-wave case with $\Phi_{a} = 0.2$ and $\Phi_{\beta {\rm min}} = 0.2$ as seen in Fig.\ \ref{fig:fig2}(b). 
We next consider another parameter set for the anisotropic $\pm s$-wave pair function below.

\begin{figure}
  \begin{center}
    \begin{tabular}{p{40mm}p{40mm}}
      \resizebox{55mm}{!}{\includegraphics{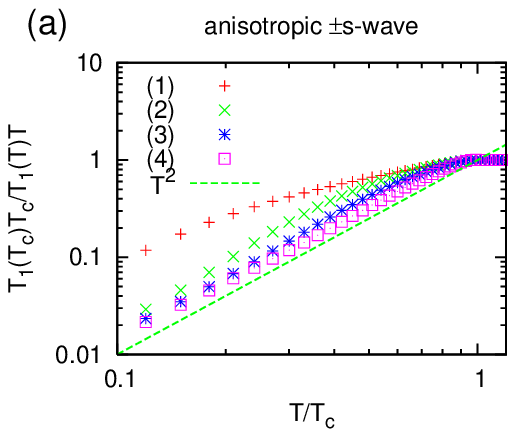}} &
      \resizebox{55mm}{!}{\includegraphics{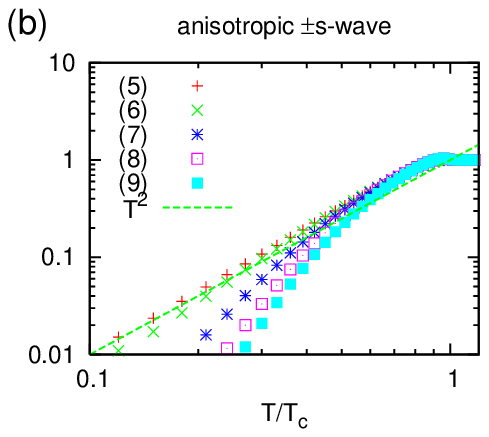}} 
    \end{tabular}
\caption{\label{fig:fig4}
(Color online) 
Temperature dependence of the nuclear magnetic relaxation rate $1/T_{1}T$ with the five band 
model. 
(a) $\Phi_{\beta {\rm min}} = 0.2$ and $\Phi_{a} = $ 0.2 (1), 0.5 (2), 0.75 (3), and 1 (4). 
(b) $\Phi_{a} = 1$ and $\Phi_{\beta {\rm min}} = $0.25 (5), 0.3 (6), 0.5 (7), 0.75 (8), and 1 (9). 
$2 \Delta_{0}/T_{\rm c} = 4$ and smearing factor $\eta = 0.1 T_{\rm c}$. 
The dashed line is a plot of $T^{2}$. }
  \end{center}
\end{figure}

\begin{figure}
\includegraphics[width = 8cm]{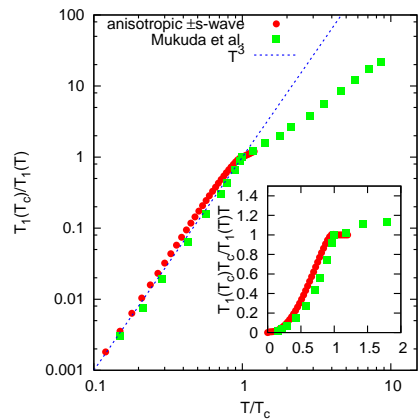}
\caption{\label{fig:fig5}
(Color online) 
Temperature dependence of the nuclear magnetic relaxation rate $1/T_{1}$ 
on a double-logarithmic scale.  
The red circles denote the result of the anisotropic $\pm s$-wave pair function 
($\Phi_{a} = 1$, $\Phi_{\beta {\rm min}} = 0.25$, 
$2 \Delta_{0}/T_{\rm c} = 4$, and smearing factor $\eta = 0.1 T_{\rm c}$). 
The green squares represent the experimental result of $^{75}$As-NQR for LaFeAsO$_{0.6}$ by Mukuda {\it et al.} (Ref.\ \onlinecite{mukuda}).
The dashed line is a plot of $T^{3}$. 
Inset: Plots of the same data for $1/T_{1}T$ on a non-logarithmic scale. 
}
\end{figure}

We search for the most suitable pair function with $\Phi_{a}$ and $\Phi_{\beta {\rm min}}$. 
We check the two points as follows: (i) the lack of the coherence peak below $T_{\rm c}$ 
and (ii) the low temperature power-law behavior $1/T_{1} T \propto T^{2}$. 
We show the temperature dependence of $1/T_{1}T$ in the cases of the various pair functions in 
Fig.\ \ref{fig:fig4}.
First, we fix the $\beta$ gap anisotropy
$\Phi_{\beta {\rm min}} = 0.2$ and examine the $\Phi_{a}$ 
(the $\alpha$ gap amplitude) dependence 
as shown in Fig.\ \ref{fig:fig4}(a). 
With increasing $\Phi_{a}$ from the value $\Phi_{a} = 0.2$, 
the exponent (i.e., the slope in Fig.\ \ref{fig:fig4}) approaches to the experimental result $\sim 2$. 
The best coincidence is attained at $\Phi_{a} = 1$. 
Second, we fix $\Phi_{a} = 1$ and examine
the $\Phi_{\beta {\rm min}}$ dependence as shown in Fig.\ \ref{fig:fig4}(b). 
With decreasing the anisotropy (i.e., increasing $\Phi_{\beta {\rm min}}$), 
the deviation becomes larger for $\Phi_{\beta {\rm min}}>0.25$. 
Hence, 
the experimental results are best reproduced when $\Phi_{a} = 1$ and $\Phi_{\beta {\rm min}} = 0.25$. 
That is, the maximum pair amplitudes 
on the Fermi surfaces $\alpha_{1,2}$ and $\beta_{1,2}$ 
are of the same order ($\Phi_{a} = 1$), and 
the ratio of the minimum to the maximum of the pair amplitude on $\beta_{1,2}$ is 0.25 ($\Phi_{\beta {\rm min}} = 0.25$).  
We show the comparison of our calculation with the experimental result of
$^{75}$As-NQR for LaFeAsO$_{0.6}$ (Ref.\ \onlinecite{mukuda}) in Fig.\ \ref{fig:fig5}. 
Indeed, this anisotropic $\pm s$-wave pair function explains the observed low-temperature 
power-law behavior $1/T_{1} \propto T^{3}$. 

\section{Superfluid density}
Let us confirm whether the above anisotropic $\pm s$-wave pair function can also explain 
the observed temperature dependence of the superfluid density. 
   The superfluid density $\rho_{xx}$ is given by\cite{choi,Haya06-1}
\begin{eqnarray}
\frac{\rho_{xx}}{ \rho_{0}}
&=&
\frac{2\pi T}
{
\Bigl\langle
      \bigl\{ v_{{\rm F} x}({\bm k}_{\rm F}) \bigr\}^2
   \Bigr\rangle_{\rm FS}
}
\sum_{\omega_n >0}
   \Biggl\langle
     \frac{
       \bigl\{ v_{{\rm F} x}({\bm k}_{\rm F}) \bigr\}^2
       \bigl|\Delta({\bm k}_{\rm F}) \bigr|^2
     }
     {
       \bigl(
         \omega_n^2 
         +\bigl| \Delta({\bm k}_{\rm F}) \bigr|^2
       \bigr)^{3/2}
     }
   \Biggr\rangle_{\rm FS}, \nonumber \\
\label{eq:rho}
\end{eqnarray} 
Here, $\rho_{0}$ denotes the superfluid density at the zero temperature and 
$v_{{\rm F} x}$ is the Fermi velocity component in the $(\pi,0)$ direction.  

\begin{figure}
\includegraphics[width = 6cm]{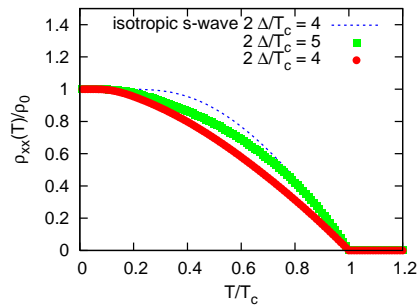}
\caption{\label{fig:fig6}
(Color online)
Temperature dependence of the superfluid density $\rho_{xx}$ for the 
anisotropic $\pm s$-wave pair function 
($\Phi_{a} = 1$, $\Phi_{\beta {\rm min}} = 0.25$). 
$2 \Delta_{0}/T_{\rm c} =$ 4 (red circles), 5 (green squares). 
The dashed line represents $\rho_{xx}(T)$ for the isotropic $s$-wave gap with $2 \Delta_{0}/T_{\rm c} = 4$. 
$\rho_{0}$ is the superfluid density at $T=0$.
}
\end{figure}

As shown in Fig.\ \ref{fig:fig6}, 
the superfluid density $\rho_{xx}(T)$ for 
the anisotropic $\pm s$-wave pair function ($\Phi_{a} = 1$, $\Phi_{\beta {\rm min}} = 0.25$) 
does not depend on the temperature in the low temperature region. 
When we increase  
$2 \Delta_{0}/T_{\rm c}$, the result 
approaches to that of the isotropic $s$-wave case.  
Indeed, the anisotropic $\pm s$-wave pair function can explain the fully-gapped behavior observed in the experiments.\cite{luetkens,khasanov,malone,hashimoto,martin,luetkens2}
In contrast to it, 
pair functions with line nodes such as $d$-wave one lead to a strong temperature dependence 
near the zero temperature in general.

\begin{figure}
\includegraphics[width = 6cm]{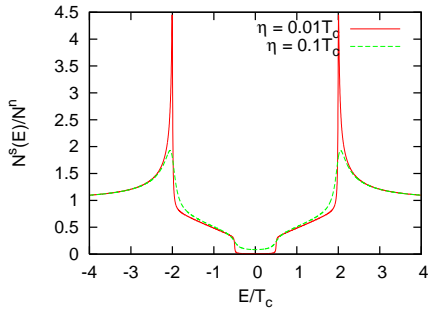}
\caption{\label{fig:fig7}
(Color online)
Energy dependence of the density of states at $T = 0$ for the anisotropic $\pm s$-wave pair function 
($\Phi_{a} = 1$, $\Phi_{\beta {\rm min}} = 0.25$, and $2 \Delta_{0}/T_{\rm c} = 4$). 
$N^{n}$ is the normal-state density of states at the Fermi level.
Plots represent the density of states with
the smearing factor $\eta=0.1 T_{\rm c}$ (green dashed line)
and $0.01 T_{\rm c}$ (red solid line).\cite{mean_free_path}
}
\end{figure}

\section{Density of states}
Finally, we show the density of states $N^{s}(E)$ for the 
anisotropic $\pm s$-wave pair function ($\Phi_{a} = 1$, $\Phi_{\beta {\rm min}} = 0.25$)
with $2 \Delta_{0}/T_{\rm c} = 4$
in Fig.\ \ref{fig:fig7}.
It is calculated by
$N^{s}(E)
=N^{n} {\rm Re} \bigl\langle g_{\uparrow\uparrow}(i\omega_n \to E+i\eta) \bigr\rangle_{\rm FS}$,\cite{mean_free_path}
where $N^{n}$ is the normal-state density of states at the Fermi level
and $g_{\uparrow\uparrow}$ is defined in Eq.\ (\ref{eq:g}).
The density of states is gapped 
in the region $|E| < \Phi_{\beta {\rm min}} \Delta_{0} =
2 \Phi_{\beta {\rm min}} T_{\rm c} =0.5 T_{\rm c}$ 
($\Phi_{\beta {\rm min}}\Delta_{0}$ is the minimum gap on the Fermi surfaces $\beta_{1,2}$). 
This is the reason why the superfluid density does not depend on the temperature in the low temperature region. 
In the region
$ \Phi_{\beta {\rm min}} \Delta_{0}(=0.5 T_{\rm c}) \alt |E| \alt \Delta_{0}(=2 T_{\rm c})$, 
the density of states has a linear energy dependence. 
Therefore, the nuclear magnetic relaxation rate exhibits the line-nodes-like power-law behavior. 
The density of states also has the single peak structure near the gap edge at $|E| = 2 T_{\rm c} = \Delta_{0}$, since 
the gap maxima on the Fermi surfaces $\alpha_{1,2}$ and $\beta_{1,2}$ now coincide with each other owing to $\Phi_{a} = 1$. 
Note here that the maximum gap amplitudes on $\alpha_{1,2}$ and $\beta_{1,2}$ are $\Delta_0 |\Phi_a|$ and $\Delta_{0}$, respectively.

In addition, 
the density of states for the anisotropic $\pm s$-wave pair function is a monotonically-increasing function of the energy ($|E|<\Delta_0$) as seen in 
Fig.\ \ref{fig:fig7}, 
while the unitary-scattering-induced density of states
and the multi-gapped density of states
are nonmonotonic in some cases.\cite{bang,preosti}
This difference would be observed by
spectroscopy experiments.

\section{Conclusion}
With the use of the five band model, 
we calculated the nuclear magnetic relaxation rate $1/T_1$ and the superfluid density $\rho_{xx}$
and showed that the anisotropic $\pm s$-wave pair function can explain the seemingly contradictory experimental results 
on Fe-pnictide superconductors. 
That is,
the anisotropic $\pm s$-wave pair function
reproduces consistently
$1/T_1 \sim T^3$ and the $T$-independence of $\rho_{xx}$ at low $T$.

Our scenario is similar to the theories by Parker {\it et al.},\cite{parker}
Chubukov {\it et al.},\cite{chubukov} and Bang and Choi\cite{bang,bang0807} in the sense 
that $\pm s$-wave pair functions are considered in all theories. 
However, impurity effects are essential for those previous theories.\cite{parker,chubukov,bang}
The impurity scattering rate is relatively large in Refs.\ \onlinecite{parker} and \onlinecite{chubukov}.
A unitary-limit impurity scattering
or an impurity scattering intermediate between Born and unitary limits\cite{preosti} is essential
in Refs.\ \onlinecite{parker} and \onlinecite{bang}.
In contrast,
we have assumed a rather clean system
and not considered a unitary-limit or an intermediate phase-shift scattering.
On the other hand,
it was pointed out that 
a fitting
resulted in
quite big value $2\Delta_0/T_{\rm c} \approx 7.5$
within a model in Ref.\ \onlinecite{bang0807}.
In our model,
rather strong gap anisotropy on the $\beta$ Fermi surfaces\cite{Arita}
 has been introduced,
which enables us to explain $1/T_1 \sim T^3$ even in a clean system
and with relatively reasonable value $2\Delta_0/T_{\rm c} \sim 4$.
This is a distinguished feature of our scenario.

It should be noted that 
while some of experimental groups have reported the fully-gapped behavior of the superfluid density, 
part of measurements showed somewhat strong temperature dependence indicating gap 
nodes.\cite{ren,luetkens,ahilan,khasanov,malone,hashimoto,luetkens2,takeshita,martin}  
Those results seem to depend on kinds of materials and doping level, but 
it is still unclear what is the essential origin of such scattered observations between materials. 
The difference might mean that the pairing symmetry changes between materials
or that the degree of gap anisotropy on the $\beta$ Fermi surfaces changes, 
albeit there are no microscopic theories suggesting them at present.
In any case,  it is an interesting issue left for feature studies.

\begin{acknowledgments}
   We thank K.~Kuroki, R.~Arita, and Y.~Kato for helpful discussions. 
   We also thank H.~Mukuda for providing us the experimental data
   (Fig.\ \ref{fig:fig5}). 
   One of us (Y.N.) acknowledges support 
   by Grand-in-Aid for JSPS Fellows (204840),
   and 
   M.M.\ is supported by JSPS Core-to-Core
   Program-Strategic Research Networks, ``Nanoscience and
   Engineering in Superconductivity (NES)".
\end{acknowledgments}

\appendix*
\section{}   
   In this Appendix,
we describe the procedure for deriving
the nuclear spin-lattice relaxation rate
$T_1^{-1}({\bm r},T)$
on the basis of
the quasiclassical Green function theory.\cite{eilen,LO,serene,kopnin,schopohl80,rieck,choi,kusunose}
The derived formula has been utilized in Sec.\ III and
in Refs.\ \onlinecite{Haya06-2,Haya03,Haya06-B,Haya03-JLTP}.

{\it Quasiclassical theory} ---
   We start with the Green functions defined as\cite{schopohl80}
\begin{subequations}
\label{eq:Green}
\begin{eqnarray}
G_{s,s'}({\bm r},{\bm r}';\tau)
=-\Bigl\langle T_\tau
\bigl[
\psi_s({\bm r},\tau) \psi_{s'}^\dagger({\bm r}',0)
\bigr]
\Bigr\rangle,
\label{eq:Green-a}
\end{eqnarray}
\begin{eqnarray}
F_{s,s'}({\bm r},{\bm r}';\tau)
=-\Bigl\langle T_\tau
\bigl[
\psi_s({\bm r},\tau) \psi_{s'}({\bm r}',0)
\bigr]
\Bigr\rangle,
\label{eq:Green-b}
\end{eqnarray}
\begin{eqnarray}
{\bar F}_{s,s'}({\bm r},{\bm r}';\tau)
=-\Bigl\langle T_\tau
\bigl[
\psi_s^\dagger({\bm r},\tau) \psi_{s'}^\dagger({\bm r}',0)
\bigr]
\Bigr\rangle,
\label{eq:Green-c}
\end{eqnarray}
\begin{eqnarray}
{\bar G}_{s,s'}({\bm r},{\bm r}';\tau)
=-\Bigl\langle T_\tau
\bigl[
\psi_s^\dagger({\bm r},\tau) \psi_{s'}({\bm r}',0)
\bigr]
\Bigr\rangle.
\label{eq:Green-d}
\end{eqnarray}
\end{subequations}
   Here,
the brackets $\langle \cdots \rangle$ denote
the thermal average.
   We use units in which $\hbar = k_{\rm B} = 1$.
   We write
\begin{equation}
{\check G}=
\begin{pmatrix}
{\hat G} &
{\hat F} \\
{\hat {\bar F}} &
{\hat {\bar G}}
\end{pmatrix}.
\label{eq:org-Green}
\end{equation}
   Throughout this Appendix,
{\it ``hat"} (${\hat A}$) denotes the $2\times2$ matrix in the spin space,
and {\it ``check"} (${\check A}$) denotes the $4\times4$ matrix
composed of the $2\times2$ particle-hole space
and the $2\times2$ spin one.

   The quasiclassical Green function ${\check g}$ is defined as
\begin{equation}
{\check g}=
{\check \tau}_3
\int d\xi_k {\check G}
\equiv
{\check \tau}_3
\begin{pmatrix}
{\hat g}^{11} &
{\hat g}^{12} \\
{\hat g}^{21} &
{\hat g}^{22}
\end{pmatrix},
\label{eq:qc-Green}
\end{equation}
where the integration is performed with respect to
the energy variable $\xi_k$ in the ${\bm k}$ space,
\begin{equation}
\xi_k \equiv
\varepsilon({\bm k})-\mu.
\label{eq:ene-xi}
\end{equation}
   Here, $\varepsilon({\bm k})$ is the quasiparticle dispersion relation
and $\mu$ is the chemical potential.
   We have defined
\begin{equation}
{\check \tau}_3 =
\begin{pmatrix}
{\hat \sigma}_0 &
0 \\
0 &
-{\hat \sigma}_0
\end{pmatrix},
\quad
{\rm with}
\quad
{\hat \sigma}_0 =
\begin{pmatrix}
1 &
0 \\
0 &
1
\end{pmatrix}.
\label{eq:tau3}
\end{equation}
   According to a conventional procedure,
the $k$-space integration is approximated as
\begin{equation}
\int \frac{d^3 k}{(2\pi)^3}
\approx
N_{\rm F} \int \frac{d\Omega}{4\pi} \int d \xi_k.
\label{eq:k-int1}
\end{equation}
Here, an isotropic spherical Fermi surface is assumed for clarity.
The extension to general cases can be done straightforward
by replacing the solid-angle integration $\int{d\Omega}/{4\pi}$ with the Fermi surface average
$\langle \cdots \rangle_{\rm FS}$.
$N_{\rm F}$ is the total density of states at the Fermi level.


   The quasiclassical Green function follows
the Eilenberger equation, which is given
as\cite{eilen,LO,serene,kopnin,schopohl80,rieck,choi,kusunose}
\begin{equation}
i {\bm v}_{\rm F} \cdot
{\bm \nabla}{\check g}
+ \bigl[ i\omega_n {\check \tau}_{3}
-{\check \Delta},
{\check g} \bigr]
=0,
\label{eq:eilen0}
\end{equation}
where
${\hat \Delta}$ is the superconducting order parameter,
\begin{equation}
{\check  \Delta} =
\begin{pmatrix}
0 &
{\hat \Delta} \\
-{\hat \Delta}^\dagger &
0
\end{pmatrix}.
\label{eq:Delta}
\end{equation}
   This equation is supplemented
by the normalization condition,\cite{eilen,schopohl80}
${\check g}^2 =
-\pi^2 {\check 1}$.

   We define, in the particle-hole space,
the matrix elements of the quasiclassical Green function ${\check g}$
as\cite{Haya06-1}
\begin{equation}
{\check g}=
-i\pi
\begin{pmatrix}
{\hat g} &
i{\hat f} \\
-i{\hat {\bar f}} &
-{\hat {\bar g}}
\end{pmatrix}.
\label{eq:qcg}
\end{equation}
Comparing Eqs.\ (\ref{eq:qc-Green}) and (\ref{eq:qcg})
we have the following relation, which we will use later.
\begin{subequations}
\label{eq:G3}
\begin{eqnarray}
{\hat g}^{11}&=&-i\pi{\hat g},  \\
{\hat g}^{22}&=&-i\pi{\hat {\bar g}},  \\
{\hat g}^{12}&=&  \pi{\hat f},  \\
{\hat g}^{21}&=&  \pi{\hat {\bar f}}.
\end{eqnarray}
\end{subequations}
In the case of spin-singlet superconductivity,
the Eilenberger equation is solved in a spatially uniform system
and the solution for
the quasiclassical Green function is\cite{Klein}
\begin{eqnarray}
{\hat g}
&=&
\frac{\omega_n {\hat \sigma}_0}{\sqrt{\omega_n^2+|\Delta|^2}},
\qquad
{\hat {\bar g} }
=
\frac{\omega_n {\hat \sigma}_0}{\sqrt{\omega_n^2+|\Delta|^2}},
\nonumber  \\
{\hat f}
&=&
\frac{\Delta i {\hat \sigma}_y }{\sqrt{\omega_n^2+|\Delta|^2}},
\qquad
{\hat {\bar f} }
=
\frac{\Delta^* (-i {\hat \sigma}_y) }{\sqrt{\omega_n^2+|\Delta|^2}}.
\end{eqnarray}
Here, the Pauli matrices are
${\hat {\bm \sigma}}=({\hat \sigma}_x,{\hat \sigma}_y,{\hat \sigma}_z)$
in the spin space.

{\it Relaxation Rate} ---
   The nuclear spin-lattice relaxation rate
$T_1^{-1}({\bm r},T)$
is obtained from the spin-spin correlation function
$\chi_{-+}(x,x')$.\cite{takigawa}
   We define
$x\equiv({\bm r}, \tau)$,
and set $\tau'=0$.
   We apply a static external magnetic field
along a certain axis
and take the spin quantization axis parallel to this.
   $\chi_{-+}(x,x')$ is given as
\begin{eqnarray}
\chi_{-+}(x,x')
&=&
\Bigl\langle T_\tau
\bigl[
S_-(x) S_+(x')
\bigr]
\Bigr\rangle
\\
&=&
\Bigl\langle T_\tau
\bigl[
\psi_\downarrow^\dagger(x) \psi_\uparrow(x)
\psi_\uparrow^\dagger(x') \psi_\downarrow(x')
\bigr]
\Bigr\rangle
\\
&=& \nonumber
{\bar G}_{\downarrow\downarrow}(x,x')
G_{\uparrow\uparrow}(x,x')
\\
& & { }
-
{\bar F}_{\downarrow\uparrow}(x,x')
F_{\uparrow\downarrow}(x,x').
\label{eq:spin-spin}
\end{eqnarray}

   Let us consider a Fourier transformation with respect to $\tau$.
   In what follows, $A$ and $B$ stand for the Green functions.
   The Fermi- and Bose-Matsubara frequencies are
$\omega_n=\pi T(2n+1)$ and $\Omega_n=\pi T(2n)$, respectively.
   The Fourier transformation is
\begin{eqnarray}
A({\bm r},{\bm r}';\tau)
=
\frac{1}{\beta} \sum_{\omega_n}
e^{-i\omega_n \tau}
A({\bm r},{\bm r}';i\omega_n).
\label{eq:fourier1}
\end{eqnarray}
   Note that
$A(\tau)$ and $A(\tau)B(\tau)$ are
periodic functions of $\tau$
with the periods $2\beta$ and $\beta$, respectively.
   Using Eq.\ (\ref{eq:fourier1}),
we have the relation
\begin{eqnarray}
\int_0^\beta d\tau e^{i\Omega_m \tau}
A(\tau)B(\tau)
&=&
\frac{1}{\beta} \sum_{\omega_n}
A(i\omega_n)B(i\Omega_m-i\omega_n).
\nonumber  \\
\label{eq:fourier2}
\end{eqnarray}
   From Eq.\ (\ref{eq:spin-spin}), the spin-spin correlation function is
\begin{eqnarray}
\chi_{-+}({\bm r},{\bm r}';i\Omega_m)
&=&
\int_0^\beta d\tau e^{i\Omega_m \tau}
\chi_{-+}({\bm r},{\bm r}';\tau)
\\
&=&     \nonumber
\int_0^\beta d\tau e^{i\Omega_m \tau}
\\ \nonumber
& & { }
\times
\Bigl[
{\bar G}_{\downarrow\downarrow}({\bm r},{\bm r}';\tau)
G_{\uparrow\uparrow}({\bm r},{\bm r}';\tau)
\\ \nonumber
& & { }
-
{\bar F}_{\downarrow\uparrow}({\bm r},{\bm r}';\tau)
F_{\uparrow\downarrow}({\bm r},{\bm r}';\tau)
\Bigr].
\\
\end{eqnarray}
   Using Eq.\ (\ref{eq:fourier2}), we obtain
\begin{eqnarray}
\label{eq:chi}
\chi_{-+}({\bm r},{\bm r}';i\Omega_m)
&=&
\frac{1}{\beta} \sum_{\omega_n} \nonumber
\\      \nonumber
& & { }
\times
\Bigl[
{\bar G}_{\downarrow\downarrow}({\bm r},{\bm r}';i\omega_n)
G_{\uparrow\uparrow}({\bm r},{\bm r}';i\Omega_m-i\omega_n)
\\ \nonumber
& & { }
-
{\bar F}_{\downarrow\uparrow}({\bm r},{\bm r}';i\omega_n)
F_{\uparrow\downarrow}({\bm r},{\bm r}';i\Omega_m-i\omega_n)
\Bigr].
\\
\end{eqnarray}

   Now, we define
${\tilde {\bm r}}\equiv{\bm r}-{\bm r}'$.
\begin{eqnarray}
A({\bm r},{\bm r}')
\equiv
A({\bm r},{\tilde {\bm r}})
=
\int \frac{d^3 k}{(2\pi)^3}
e^{i{\bm k}\cdot {\tilde {\bm r}} }
A({\bm r},{\bm k}).
\end{eqnarray}
   Setting ${\bm r}'={\bm r}$ (i.e., ${\tilde {\bm r}}=0$), we have
\begin{eqnarray}
A({\bm r},{\bm r})
&=&
\int \frac{d^3 k}{(2\pi)^3}
A({\bm r},{\bm k})
\\
&\approx&
N_{\mathrm F} \int \frac{d\Omega}{4\pi}
\int d\xi_k
A({\bm r};{\bar {\bm k}},\xi_k).
\label{eq:2.4}
\end{eqnarray}
   Here,
we have referred to Eq.\ (\ref{eq:k-int1}) (quasiclassical approximation).
   ${\bar {\bm k}}$ denotes the position of ${\bm k}$ on the Fermi surface.
   Then, from Eqs.\ (\ref{eq:chi}) and (\ref{eq:2.4}),
$\chi_{-+}({\bm r},{\bm r}'={\bm r};i\Omega_m)$ is
\begin{eqnarray}
& & \nonumber
\chi_{-+}({\bm r},{\bm r};i\Omega_m)
\\ \nonumber
&=&
N_{\mathrm F}^2
\frac{1}{\beta} \sum_{\omega_n}
\\      \nonumber
& & { } \times
\Biggl[
\Bigl\langle
g^{22}_{\downarrow\downarrow}({\bm r},{\bar {\bm k}};i\omega_n)
\Bigr\rangle_{\rm FS}
\Bigl\langle
g^{11}_{\uparrow\uparrow}({\bm r},{\bar {\bm k}};i\Omega_m-i\omega_n)
\Bigr\rangle_{\rm FS}
\\  \nonumber
& & {}
\qquad -
\Bigl\langle
g^{21}_{\downarrow\uparrow}({\bm r},{\bar {\bm k}};i\omega_n)
\Bigr\rangle_{\rm FS}
\Bigl\langle
g^{12}_{\uparrow\downarrow}({\bm r},{\bar {\bm k}};i\Omega_m-i\omega_n)
\Bigr\rangle_{\rm FS}
\Biggr].
\label{eq:2.18}
\\
\end{eqnarray}
Here,
we have referred to Eq.\ (\ref{eq:qc-Green})
and have replaced $\int {d\Omega}/{4\pi}$
with $\langle \cdots \rangle_{\rm FS}$.

   Next, let us consider
the spectral representation of the quasiclassical Green functions:
\begin{eqnarray}
{\hat A}(i\omega_n)=\int_{-\infty}^{\infty}
d\omega
\frac{{\hat a}(\omega)}{i\omega_n-\omega}.
\end{eqnarray}
   Utilizing the formula ($f(\omega)$ is the Fermi distribution function)
\begin{eqnarray}
\frac{1}{\beta} \sum_{\omega_n}
\frac{1}{(i\omega_n-\omega)(i\Omega_m-i\omega_n-\omega')}
=
\frac{f(-\omega')-f(\omega)}{\omega+\omega'-i\Omega_m},
\nonumber
\\
\end{eqnarray}
we calculate
\begin{eqnarray}
Q(i\Omega_m)
&\equiv&
\frac{1}{\beta} \sum_{\omega_n}
A(i\omega_n)B(i\Omega_m-i\omega_n)
\label{eq:2.23}
\\
&=&
\int_{-\infty}^{\infty}d\omega
\int_{-\infty}^{\infty}d\omega'
a^A(\omega)a^B(-\omega')
\frac{f(\omega')-f(\omega)}{\omega-\omega'-i\Omega_m}.
\nonumber \\
\end{eqnarray}
   Setting $i\Omega_m \rightarrow \Omega + i\delta$
($\delta \rightarrow 0^+$),
\begin{eqnarray}
Q(\Omega)
&=&
\int_{-\infty}^{\infty}d\omega
\int_{-\infty}^{\infty}d\omega'
a^A(\omega)a^B(-\omega')
\frac{f(\omega')-f(\omega)}{\omega-\omega'-\Omega-i\delta}
\nonumber \\
\\
&=&   \nonumber
{\bm {\mathrm P}} \int_{-\infty}^{\infty}d\omega
\int_{-\infty}^{\infty}d\omega'
a^A(\omega)a^B(-\omega')
\frac{f(\omega')-f(\omega)}{\omega-\omega'-\Omega}
\\  \nonumber
& & { }
+i\pi
\int_{-\infty}^{\infty}d\omega
\int_{-\infty}^{\infty}d\omega'
\\ \nonumber
& & { }
\times
a^A(\omega)a^B(-\omega')
\bigl\{f(\omega')-f(\omega)\bigr\}
\delta(\omega-\omega'-\Omega),
\\
\end{eqnarray}
where we have used
\begin{eqnarray}
\frac{1}{\omega-\omega'-\Omega\pm i\delta}
=
{\bm {\mathrm P}}
\frac{1}{\omega-\omega'-\Omega}
\mp i\pi
\delta(\omega-\omega'-\Omega).
\nonumber \\
\label{eq:2.31}
\end{eqnarray}
   Thus,
\begin{eqnarray}
\lim_{\Omega\rightarrow 0^+} {\rm Im} \frac{Q(\Omega)}{\Omega}
&=&
\label{eq:2.35}
\frac{\pi\beta}{4}
\int_{-\infty}^{\infty} d\omega
a^A(\omega)a^B(-\omega)
\frac{1}{\cosh^2(\beta\omega/2)}.
\nonumber \\
\end{eqnarray}

   It is known that
$T_1^{-1}({\bm r},T)$ is calculated by\cite{takigawa} ($\delta \rightarrow 0^+$)
\begin{eqnarray}
\label{eq:T1-T1}
T_1^{-1}({\bm r},T)
=
T
\lim_{\Omega\rightarrow 0^+} {\rm Im}
\frac{\chi_{-+}
({\bm r},{\bm r};i\Omega_m \rightarrow \Omega + i\delta)}{\Omega}.
\nonumber \\
\end{eqnarray}
   Referring to Eqs.\ (\ref{eq:2.18}), (\ref{eq:2.23}), (\ref{eq:2.35}), and (\ref{eq:T1-T1}),
we obtain
\begin{eqnarray}
T_1^{-1}({\bm r},T)
&=&   \nonumber
\frac{\pi N_{\mathrm F}^2}{4}
\int_{-\infty}^{\infty} d\omega
\frac{1}{\cosh^2(\omega/2T)}
\\ \nonumber
& & { }
\times
\Bigl[
\bigl\langle a^{22}_{\downarrow\downarrow}(\omega) \bigr\rangle_{\rm FS}
\bigl\langle a^{11}_{\uparrow\uparrow}(-\omega) \bigr\rangle_{\rm FS}
\\ \nonumber
& & { }
-
\bigl\langle a^{21}_{\downarrow\uparrow}(\omega) \bigr\rangle_{\rm FS}
\bigl\langle a^{12}_{\uparrow\downarrow}(-\omega) \bigr\rangle_{\rm FS}
\Bigr].
\\
\end{eqnarray}
    In the normal state,
the spectral function of the quasiclassical Green function
is $a^{11}=a^{22}=1$
for diagonal components in the particle-hole space
 (i.e., the density of states is unity in units of $N_{\mathrm F}$)
and is $a^{12}=a^{21}=0$ for off-diagonal components
(because the order parameter is zero).
   We then obtain at $T=T_{\mathrm c}$,
\begin{eqnarray}
T_1^{-1}({\bm r},T_{\mathrm c})
&=&
\frac{\pi N_{\mathrm F}^2}{4}
\int_{-\infty}^{\infty} d\omega
\frac{1}{\cosh^2(\omega/2T_{\mathrm c})}
\\
&=&
\pi T_{\mathrm c} N_{\mathrm F}^2.
\end{eqnarray}
Hence, the relaxation rate presented in Sec.\ III is obtained:
\begin{eqnarray}
\frac{T_1({\bm r},T_{\mathrm c})T_{\mathrm c}}{T_1({\bm r},T)T}
&=&   \nonumber
\frac{1}{4T }
\int_{-\infty}^{\infty} d\omega
\frac{1}{\cosh^2(\omega/2T)}
\\ \nonumber
& & { }
\times
\Bigl[
\bigl\langle a^{22}_{\downarrow\downarrow}(\omega) \bigr\rangle_{\rm FS}
\bigl\langle a^{11}_{\uparrow\uparrow}(-\omega) \bigr\rangle_{\rm FS}
\\ \nonumber
& & { }
-
\bigl\langle a^{21}_{\downarrow\uparrow}(\omega) \bigr\rangle_{\rm FS}
\bigl\langle a^{12}_{\uparrow\downarrow}(-\omega) \bigr\rangle_{\rm FS}
\Bigr].
\\
\label{eq:t1t}
\end{eqnarray}

{\it Spectral functions} ---
   In the spectral representation,
the quasiclassical Green functions are
\begin{eqnarray}
{\hat g}^{ij}(i\omega_n)=\int_{-\infty}^{\infty}
d\omega
\frac{{\hat a}^{ij}(\omega)}{i\omega_n-\omega},
\end{eqnarray}
where $i,j=\{1,2\}$.
   Letting $i\omega_n \rightarrow E\pm i\eta$ ($\eta>0$),
\begin{eqnarray}
{\hat g}^{ij}(i\omega_n\rightarrow E\pm i\eta)
&=&
\int_{-\infty}^{\infty}
d\omega
\frac{{\hat a}^{ij}(\omega)}{E-\omega\pm i\eta}
\\
&=&
{\mathrm {\bf P}}
\int_{-\infty}^{\infty}
d\omega
\frac{{\hat a}^{ij}(\omega)}{E-\omega}
\mp i\pi
{\hat a}^{ij}(E),
\nonumber \\
\end{eqnarray}
where Eq.\ (\ref{eq:2.31}) has been used.
   From this, we have the relation
\begin{eqnarray}
{\hat a}^{ij}({\bm r},{\bar {\bm k}},E)
&=&
\frac{i}{2\pi}
\Bigl[
{\hat g}^{ij}({\bm r},{\bar {\bm k}},i\omega_n\rightarrow E+ i\eta)
\nonumber \\
& & { }
-
{\hat g}^{ij}({\bm r},{\bar {\bm k}},i\omega_n\rightarrow E- i\eta)
\Bigr].
\end{eqnarray}
   Referring to Eq.\ (\ref{eq:G3}),
we have
\begin{eqnarray}
{\hat a}^{11}({\bm r},{\bar {\bm k}},E)
&=&
\frac{1}{2}
\Bigl[
{\hat g}({\bm r},{\bar {\bm k}},i\omega_n\rightarrow E+ i\eta)
\nonumber \\
& & { }
-
{\hat g}({\bm r},{\bar {\bm k}},i\omega_n\rightarrow E- i\eta)
\Bigr],
\\
{\hat a}^{22}({\bm r},{\bar {\bm k}},E)
&=&
\frac{1}{2}
\Bigl[
{\hat {\bar g}}({\bm r},{\bar {\bm k}},i\omega_n\rightarrow E+ i\eta)
\nonumber \\
& & { }
-
{\hat {\bar g}}({\bm r},{\bar {\bm k}},i\omega_n\rightarrow E- i\eta)
\Bigr],
\\
{\hat a}^{12}({\bm r},{\bar {\bm k}},E)
&=&
\frac{i}{2}
\Bigl[
{\hat f}({\bm r},{\bar {\bm k}},i\omega_n\rightarrow E+ i\eta)
\nonumber \\
& & { }
-
{\hat f}({\bm r},{\bar {\bm k}},i\omega_n\rightarrow E- i\eta)
\Bigr],
\\
{\hat a}^{21}({\bm r},{\bar {\bm k}},E)
&=&
\frac{i}{2}
\Bigl[
{\hat {\bar f}}({\bm r},{\bar {\bm k}},i\omega_n\rightarrow E+ i\eta)
\nonumber \\
& & { }
-
{\hat {\bar f}}({\bm r},{\bar {\bm k}},i\omega_n\rightarrow E- i\eta)
\Bigr].
\end{eqnarray}
   To calculate the relaxation rate in Eq.\ (\ref{eq:t1t}),
we need to consider the spin-space matrix elements presented in Sec.\ III.
%




\begin{thebibliography}{00}
\bibitem{mean_free_path}
The mean free path $l$ is estimated as
$l/\xi_0 = \pi\Delta_0/2\eta$,
provided $l=v_{\rm F} \tau$, the scattering rate $\eta=1/2\tau$, and
the zero-temperature coherence length $\xi_0=v_{\rm F}/\pi\Delta_0$.
When $\eta=0.1T_{\rm c}$ and $2\Delta_0/T_{\rm c}=4$,
the mean free path is $l \approx 30 \xi_0$
and the system is rather clean.
\end{thebibliography}
\end{document}